\documentclass[10pt,journal,twocolumn,twoside]{IEEEtran} 
\usepackage{xpatch}
\usepackage{graphicx}
\usepackage{caption}
\usepackage{epstopdf}
\usepackage{float}
\usepackage{algorithmic}
\usepackage{array}
\usepackage{amsmath}
\usepackage{amssymb}
\usepackage{mdwmath}
\usepackage{mdwtab}
\usepackage{eqparbox}
\usepackage{stfloats}
\usepackage{fixltx2e}
\usepackage{cases} 
\usepackage{flushend}
\usepackage{threeparttable}
\usepackage{xcolor}
\usepackage{makecell}
\usepackage{etoolbox}
\makeatletter
\patchcmd{\@makecaption}
  {\scshape}
  {}
  {}
  {}
\makeatletter
\patchcmd{\@makecaption}
  {\\}
  {.\ }
  {}
  {}
\makeatother
\def\figurename{FIGURE}

\usepackage[boxed,ruled,commentsnumbered]{algorithm2e}
\usepackage{url}
\usepackage{cite}
\ifCLASSOPTIONcompsoc
\usepackage[caption=false,font=normalsize,labelfont=sf,textfont=sf]{subfig}
\else
\usepackage[caption=false,font=footnotesize]{subfig}
\fi
\allowdisplaybreaks[4]

\makeatletter

\newcommand{\Rmnum}[1]{\expandafter\@slowromancap\romannumeral #1@}
\makeatother

\makeatletter
\renewcommand*{\@opargbegintheorem}[3]{\trivlist
      \item[\hskip \labelsep{\bfseries #1\ #2}] \textbf{(#3):}\ }
\makeatother

\begin{document}

\makeatletter
\def\changeBibColor#1{%
  \in@{#1}{}
  \ifin@\color{blue}\else\normalcolor\fi
}
 
\xpatchcmd\@bibitem
  {\item}
  {\changeBibColor{#1}\item}
  {}{\fail}
 
\xpatchcmd\@lbibitem
  {\item}
  {\changeBibColor{#2}\item}
  {}{\fail}
\makeatother

\title
{Privacy and Security in Ubiquitous Integrated Sensing and Communication: Threats, Challenges and Future Directions}
\author{Kaiqian Qu,  Jia Ye,~\IEEEmembership{Member, IEEE,} Xuran Li, 
and Shuaishuai Guo,~\IEEEmembership{Senior Member, IEEE}
\thanks{The work is supported in part by the National Natural Science Foundation of China under Grant 62171262 and Grant 62301090; in part by Shandong Provincial Natural Science Foundation under Grant ZR2021YQ47, ZR2021LZHP003 and ZR2023QF015; in part by the Taishan Young Scholar under Grant tsqn201909043; in part by Major Scientific and Technological Innovation Project of Shandong Province under Grant 2020CXGC010109.}
\thanks{K. Qu and S. Guo  are with the School of Control Science and Engineering, Shandong University, and also with Shandong Key Laboratory of Wireless Communication Technologies, Shandong University, China (e-mail: qukaiqian@mail.sdu.edu.cn; shuaishuai$\_$guo@sdu.edu.cn). }
\thanks{Jia Ye is with the School of Electrical Engineering, Chongqing University, Chongqing, 400044, China (yejiaft@163.com).}

\thanks{Xuran Li is with Shandong Key Laboratory of Medical Physics and Image Processing, School of Physics and Electronics, Shandong Normal University, China (e-mail: sdnulxr@sdnu.edu.cn).}
   }
\maketitle


\begin{abstract} 
 Integrated sensing and communication (ISAC) technology is at the forefront of next-generation communication, enhancing applications from intelligent transportation to unmanned aerial vehicle surveillance and healthcare through ubiquitous sensing capability. However, the advent of ISAC brings with it a dual-edged challenge. While it opens up new avenues for application, it also raises significant concerns regarding the privacy of sensitive data and the security of systems in an inherently open environment. To navigate these challenges and unlock ISAC's full potential, this article starts with an analysis of the underlying factors that contribute to privacy and security vulnerabilities. It also identifies the technical obstacles that stand in the way of achieving a secure and private ISAC ecosystem within the confines of conventional network frameworks.  
As a solution, we introduce a security and privacy-preserving network (SPPN), designed to mitigate these potential threats and tackle the technical challenges through a secure framework for information handling and collaborative efforts among trusted parties, ISAC operators, and users. Central to SPPN's security strategies is an artificial intelligence based system, equipped to detect unauthorized access and coordinate defense mechanisms, including deploying friendly jammers to thwart malicious devices and utilizing reconfigurable intelligent surfaces  (RISs) to bolster the network's defense mechanisms against potential attacks. Through detailed analysis and case studies, this paper demonstrates the effectiveness of the SPPN framework, which not only strengthens the network's defenses but also significantly improves its sensing and communication capabilities.
\end{abstract}


\section{Introduction} 
\IEEEPARstart{W}{ith} the advancement of the sixth generation (6G) wireless communication research, many potential scenarios that cannot be fully realized by the fifth generation (5G) wireless communication are proposed, such as digital twins, autonomous driving, etc. These emerging applications require the collection and transmission of massive data. In order to avoid the expensive cost of a large number of sensors, it is an innovative method to use electromagnetic waves for communication transmission while performing sensing functions to collect data. Thus, a paradigm shift  from traditional communications-based networks to integrated sensing and communication (ISAC) networks has become the key to 6G enabling emerging applications\cite{9737357}.

Sensing and communication are major consumers of wireless spectrum that are facing resource scarcity. It improves both the spectral and energy efficiency to share the same spectrum and power between communication and radar. In ISAC, the waveform conducts the communication tasks in the forward channel and sensing tasks in the backward channels. Figuring out the fundamental performance trade-off is essential. Recently, \cite{10217179} has explored the communication perception performance boundary of distributed ISAC networks. Besides, the joint signal design also attracts a lot of attention.  Recently, a beamforming strategy of vehicular-mounted ISAC units based on vehicle state was studied in \cite{10063187}; Cong \emph{et al.} proposed a general ISAC beamforming design for systems with the finite alphabet modulation set as input \cite{CongFinite2022}. Guo \emph{et al.} proposed a beamspace modulation strategy to improve the communication capacity while maintaining similar sensing performance \cite{ShuaishuaiGuo:Mobicom2022}. 

The prerequisite to deploying ISAC base stations is that their security must be guaranteed. In traditional communications, due to the broadcast characteristics of wireless channels, the transmitted signal is exposed to vulnerable environments and is easily wiretapped by the malicious eavesdropper \cite{8525328}. As a
complement to cryptography, physical layer security (PLS) is
proposed to safeguard private information from eavesdropping. PLS is capable of exploiting the physical characteristics of wireless channels, e.g., interference, fading, noise, directivity, and disparity, without introducing complicated secret key generation and management. ISAC, as the new paradigm for future wireless networks, deserves the same security concerns. 
However, only a small portion of the literature has examined issues related to ISAC security.  \cite{9199556,9737364, 10143420, yang2022secure} considered the scenario in which the eavesdropping target and the communication user are present at the same time. In order to avoid communication data leakage, they proposed to use reconfigurable intelligent surfaces (RISs), precoding/beamforming techniques, or artificial jamming signals to reduce the signal-to-intereference-plus-noise ratio (SINR) at eavesdropping targets. \cite{DongdongLi} proposed a securing ISAC framework using aerial RIS against unfriendly jamming signals.

Much of the above literature on ISAC security primarily focuses on the risk of data interception by eavesdroppers, often neglecting the significant issues of user privacy and sensing data vulnerabilities within ISAC networks. These networks, powered by omnipresent electromagnetic waves, have the capability to gather extensive data, exemplified by technologies that enable human behavior sensing through WiFi signals or breath detection via user equipment, as noted in studies like \cite{8897594}. With the incorporation of unmanned aerial vehicles (UAVs), ISAC's reach and the variety of detectable scenarios are set to widen, transcending traditional limitations faced by camera-based systems, such as line-of-sight (LoS) and ambient light requirements, through advanced methods like radar imaging.

This expansion, however, ushers in profound privacy concerns. Electromagnetic waves,  less fettered by physical barriers, can inadvertently expose sensitive personal behaviors to surveillance, a prospect most individuals find unsettling. The scenario reveals a critical balance that ISAC networks must navigate: enhancing sensing capabilities for greater safety and utility while safeguarding personal privacy. The intrusion of electromagnetic waves into private spaces without consent underlines the pressing need for robust privacy protection mechanisms in ISAC implementations. Addressing this dilemma requires a holistic strategy that marries technological innovation with policy reform and ethical guidance, ensuring ISAC networks' advantages do not erode fundamental privacy rights.
\color{black}
\section{Privacy and Security Threats}
\begin{figure*}[htbp]
       \centering
        \includegraphics[width=0.95\linewidth]{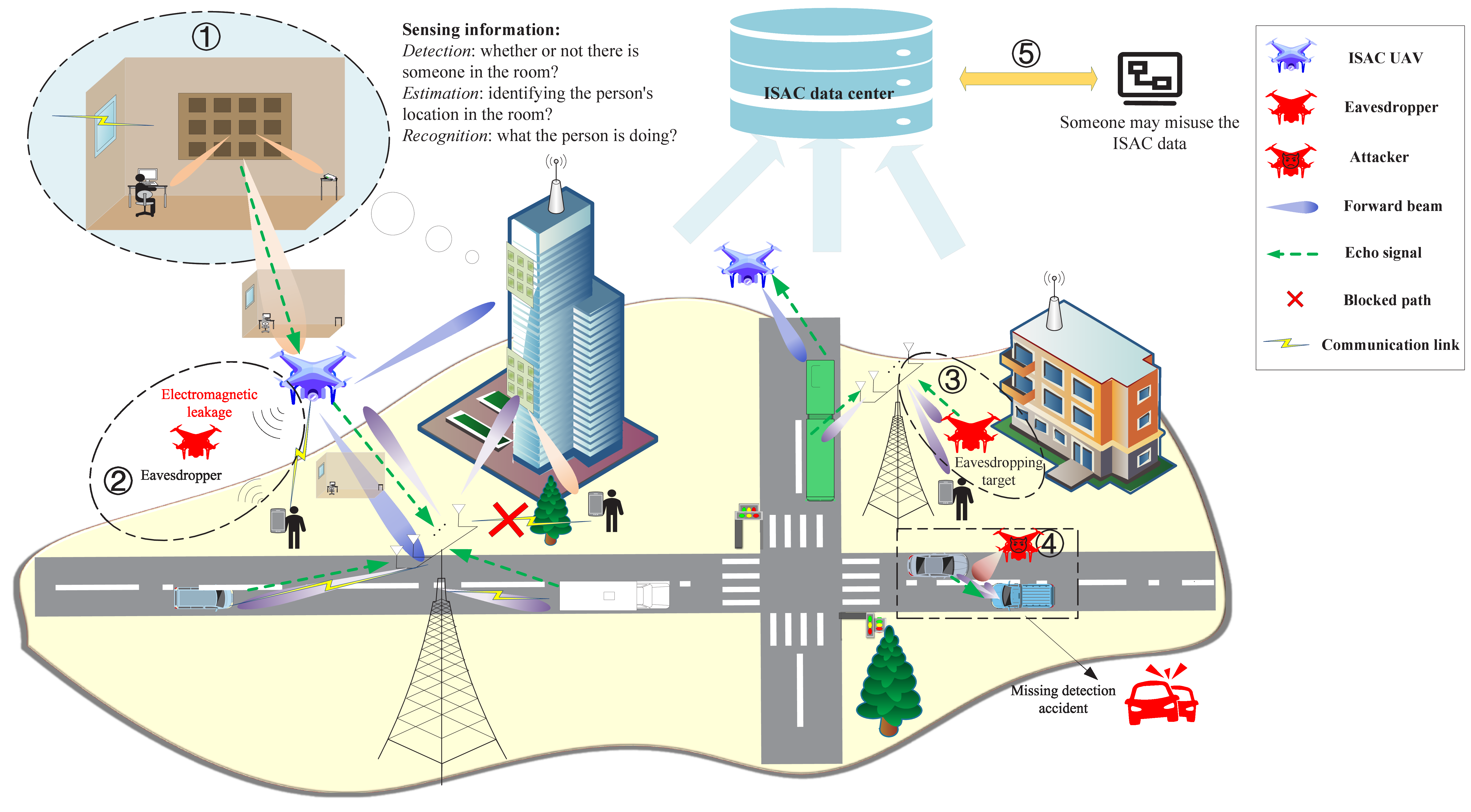}
       \caption{Potential ISAC privacy and security threats: \textcircled{1} Privacy threats from ubiquitous electromagnetic sensing; \textcircled{2}
       Security threats posed by electromagnetic leakage and
eavesdropping devices; \textcircled{3} Security threats posed by eavesdropping targets; \textcircled{4} Security threats from potential attackers; and \textcircled{5}Security and privacy issues in the ISAC data center.}
       \label{Model}
\end{figure*}
In this section, we discuss the privacy and security threats of ISAC in detail.

\subsubsection{User Privacy Threats posed by Ubiquitous Electromagnetic Sensing}
The pervasive integration of electromagnetic sensing in ISAC systems heralds significant privacy implications. Within ISAC systems, the capabilities of sensing are organized into three progressive tiers:
(1) Detection, which ascertains the presence or absence of a target;
(2) Estimation, which gauges a known target's location, velocity, and additional metrics;
(3) Recognition, which captures intricate details pertaining to a target's identity, attributes, and condition.
Particularly concerning is when these technologies are applied to monitor individuals. They enable the acquisition of sensitive information about a person's existence, movements, and activities, as exemplified in the first scenario illustrated in Fig. \ref{Model}. This level of surveillance, mirroring the functionalities of traditional camera systems yet far surpassing them in intrusiveness, employs electromagnetic waves capable of penetrating the cover of darkness and physical barriers. This allows for monitoring under conditions where optical surveillance systems fall short, raising urgent questions about privacy rights in the age of ubiquitous electromagnetic sensing.

\subsubsection{Security Threats posed by Electromagnetic Leakage and Eavesdropping Devices}
The inherent nature of wireless communication channels renders the ISAC signals particularly vulnerable to interception. This susceptibility is exacerbated by the ubiquitous risk of electromagnetic leakage, making the signals an easy target for eavesdropping devices. As illustrated in the second scenario of Fig. 1, such wiretapping activities can compromise the integrity of ISAC systems. By intercepting ISAC signals, unauthorized entities gain the potential to access sensitive communication and sensing data intended for legitimate users, posing a significant threat to the security and privacy of the system. 

\subsubsection{Security Threats posed by Eavesdropping Targets}

In the context of ISAC systems, the base station emits an integrated signal that serves dual purposes: conveying valuable information to communication users and detecting targets. Traditional radar operations necessitate concentrating the transmission power in specific directions to accurately estimate target locations and characteristics. However, this approach introduces a significant vulnerability when the 'target' is, in fact, a malicious eavesdropper. Such adversaries can exploit this focused transmission to intercept critical communication content embedded within the integrated signal, as depicted in the third scenario of Fig. 1. This scenario outlines a distinct security threat, diverging from the conventional risks of electromagnetic leakage found in traditional communication systems. The deliberate direction of signals towards potential threats, aimed at enhancing estimation accuracy, inadvertently facilitates the theft of sensitive information. Consequently, it becomes crucial to devise strategies that prevent the disclosure of communication data to malicious entities, all while maintaining effective detection capabilities. This delicate balancing act calls for innovative solutions that safeguard against information leakage to eavesdropping targets, as underscored by research in \cite{10143420,9199556,9737364}.

\subsubsection{Security Threats from Potential Attackers}
The risk of potential attackers poses another significant security challenge, particularly with regard to the accuracy of sensing operations. These malicious activities can directly compromise the safety and reliability of devices and the broader systems they integrate with. For instance, inaccuracies induced in the sensing mechanisms of ISAC-enabled vehicle networks could precipitate serious accidents, a situation depicted in the fourth scenario of Fig. 1. The degree of threat varies across different applications of ISAC, making the precision of sensing a critical factor in ensuring operational safety. Consequently, it is imperative to tailor the design of ISAC systems to meet specific quality-of-service (QoS) requirements, adapting the level of sensing accuracy to mitigate the risks posed by these security threats effectively.
\color{black}

\subsubsection{Security and Privacy Issues in the ISAC Data Center}

The potential for misuse of data presents significant concerns within ISAC systems. Should the data captured by ISAC technologies fall into unauthorized hands, it risks being exploited in ways that could lead to severe privacy infringements and harm, as illustrated in the fifth scenario of Fig. 1. Such misuse could range from unauthorized surveillance activities to the unwarranted collection of sensitive information, whether by governmental bodies or private entities. The deployment of radar sensing for surveillance purposes without appropriate legal authorization or oversight raises serious questions about the impact on individual privacy and civil liberties. It underscores the critical need for establishing robust legal frameworks and ethical guidelines to govern the use of ISAC technologies, ensuring that surveillance practices, if conducted, are carried out with transparency, accountability, and respect for individual rights.

\section{Technical Challenges to Implement Privacy-Preserving and Secure ISAC Networks}
Given the privacy and security concerns raised by ISAC usage, developing networks that prioritize these aspects is crucial. While numerous techniques—like authentication, physical layer security, encryption, covert transmission, and secure data aggregation—are known for enhancing privacy and security, ISAC's dual nature of sensing and communication poses a unique challenge. It's essential to evaluate if traditional methods for radar and wireless systems are suitable for ISAC. Furthermore, the focus has largely been on securing sensing and communication, with less attention to safeguarding individual privacy. This section explores these challenges from the perspectives of privacy preservation and security.

\subsubsection{Lack of User Awareness}
ISAC networks, unlike traditional cameras or physical sensors with visible presence, operate covertly, often unbeknownst to users. This invisibility means individuals may be unaware of being monitored or how their data is collected. ISAC employs passive sensing, where devices silently detect changes in the wireless environment—like WiFi sensing technologies tracking movements without direct user interaction. This mode of data collection obscures awareness, leaving users oblivious to the information being gathered about them. Although passive sensing boosts efficiency and reduces energy use, it diminishes user control over personal data collection. Moreover, ISAC systems in public or commercial spaces seldom inform individuals of their operation, exacerbating the issue of user awareness and control. 
\subsubsection{Unknown and Uncontrollable Wireless Environment}
 ISAC networks, utilizing electromagnetic waves, can breach privacy by penetrating walls and operating without user consent, creating an uncontrollable and unpredictable environment. This complexity hinders users from identifying or implementing effective privacy measures. Accessing propagation channel information or locating sensing devices is challenging, making it difficult for individuals to understand the extent of monitoring zones. Consequently, users may unknowingly enter areas under surveillance, amplifying privacy concerns. While blocking or interfering with sensing links is conceivable, it requires detailed channel information and incurs extra resource use and complex design challenges. The nature of this environment facilitates the collection and amalgamation of data from varied sources, heightening the risk of inference attacks. Even non-sensitive data, when pooled together, can reveal personal details, introducing unforeseen privacy threats.
\subsubsection{Lack of Trusted Management Organization}
Without a trusted management organization, users lack the recourse to opt out of ISAC's detection, estimation, or recognition processes, leaving their privacy unprotected. Such an organization is crucial for enforcing privacy protections to shield sensitive information from unauthorized access. Its absence raises concerns over data handling, storage, and potential misuse. Additionally, as ISAC networks may share data with third parties or collaborate, a trusted intermediary ensures that any data exchange respects privacy and complies with regulations. This entity is also charged with securing user consent for data collection and maintaining transparency about data use. Essentially, a trusted organization guarantees that user data is managed responsibly and aligned with user expectations, providing a layer of security and trust essential for privacy assurance.

\subsubsection{Trade-off Between Sensing and Privacy}
In ISAC networks, the act of sensing embodies a paradox, serving both as a tool for gathering valuable information and a potential threat to privacy. Sensing, by its nature, involves the collection of data about individuals or targets. For legitimate users, the accumulation of data heightens the risk of privacy breaches—the more data collected, the greater the vulnerability. As such, there's a critical balance to strike: collecting necessary information for legitimate sensing purposes while ensuring it does not infringe on user privacy, often requiring explicit user consent. Conversely, when dealing with illegitimate users, such as eavesdroppers or attackers, the goal shifts towards gathering as much information as possible to mitigate threats, all while safeguarding the communication data of legitimate users from leakage and minimizing the impact of malicious activities. This scenario presents a complex trade-off: excessive data collection from legitimate users infringes on their privacy, yet insufficient monitoring of malicious entities compromises their security. Navigating this balance between effective sensing and robust privacy protection represents a significant challenge within ISAC networks.

\subsubsection{Information-Carrying Sensing Signals}
The blend of sensing and communication in ISAC networks means that not only legitimate receivers but also unauthorized entities might intercept privacy-sensitive information. Upper-layer encryption can hinder malicious decoding attempts, yet it demands more from channel resources due to the complexities of encryption algorithms, key management, and exchange processes. Furthermore, the high computational power of potential eavesdroppers cannot be overlooked, posing a risk to information security. For sensing-focused ISAC systems, prioritizing accurate data collection over complex communication modulations presents a dilemma. Applying encryption to sensing signals without affecting their quality or the system's detection efficiency is a significant challenge, highlighting the need for balanced solutions that protect privacy without compromising sensing effectiveness.

\subsubsection{Resource Constraints}
ISAC networks often operate within the constraints of environments like wireless sensor networks or IoT devices, which are limited in computational power, memory, and energy. Balancing the dual functions of sensing and communication for system optimization in these settings presents significant challenges. The addition of privacy and security measures further complicates this, turning a dual-optimization problem into one that also needs to account for resource limitations. To address these complexities, the development of lightweight beamforming algorithms, efficient privacy and security protocols, and optimized resource management is crucial. These strategies aim to reduce computational and energy demands, enabling effective ISAC operation within resource-constrained environments.

In conclusion, establishing privacy-focused and secure ISAC networks poses distinct challenges. Addressing these is crucial for crafting resilient ISAC systems that protect sensitive data and ensure seamless performance in complex environments. Next, we will explore potential solutions and case studies to enhance ISAC networks' privacy and security.
\begin{figure*}[ht]
\centering
\includegraphics[width=0.95\linewidth,height=0.52\linewidth]{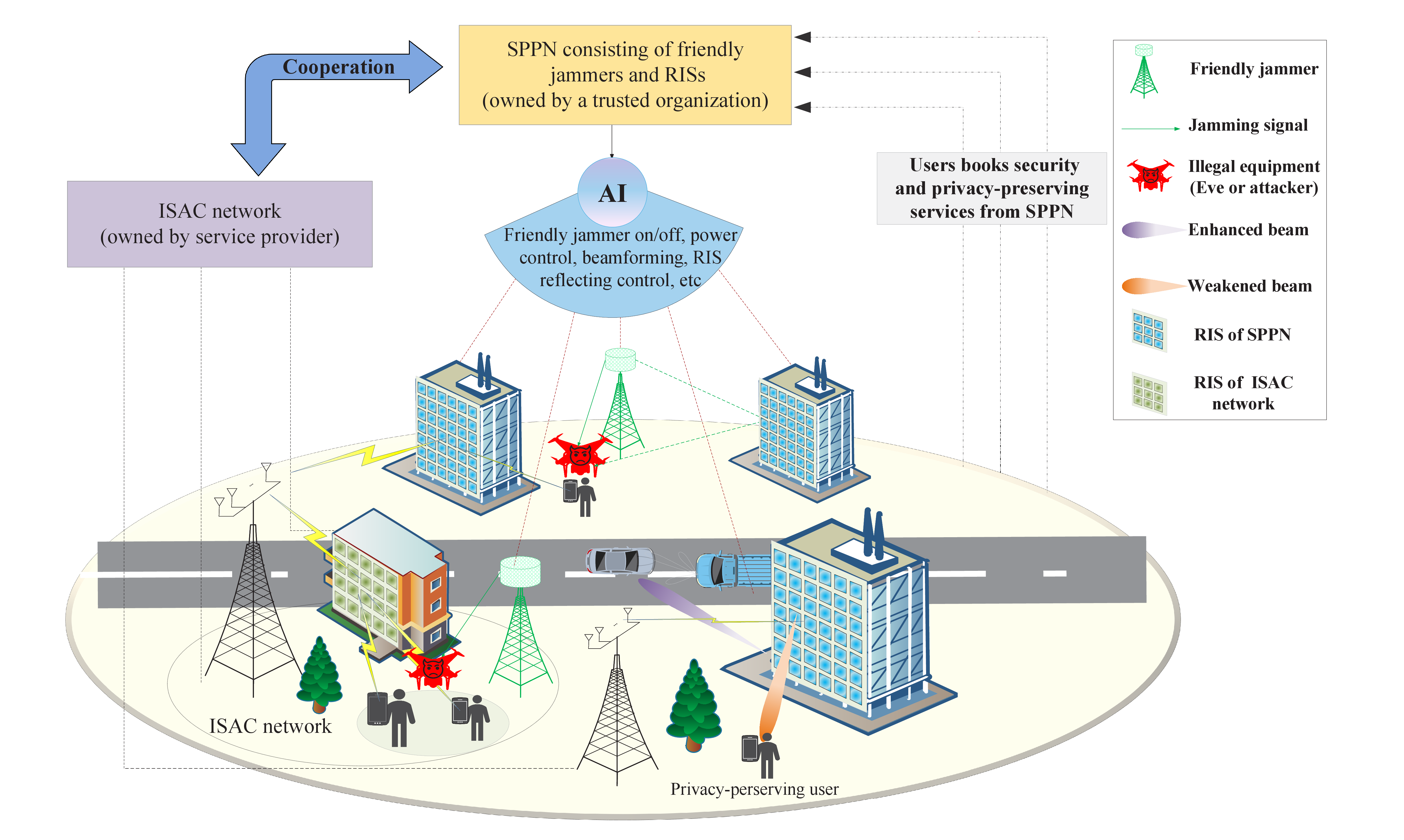}
\caption{Introducing security and privacy-preserving network (SPPN) to enhance data security and privacy within ISAC networks.}
\label{fig: SPPN}
\end{figure*}
\section{Promising Solutions and Future Directions} 

In this section, we introduce a security and privacy-preserving network (SPPN) to enhance data security and privacy within ISAC networks, as illustrated in Fig. \ref{fig: SPPN}. Drawing inspiration from storing valuables in a secure, government-regulated facility, SPPN combines trusted management with advanced technological safeguards. Key to preventing data theft are AI-based decision systems for threat analysis and additional layers of defense through friendly jammers and RISs, ensuring comprehensive protection.

The SPPN framework is divided into two core areas: data management involving trusted organizations and ISAC operations, and security enforcement through AI, jammers, and RISs. This setup not only strengthens defenses but also boosts sensing and communication effectiveness, with AI refining decision-making processes through continuous data analysis.

However, the emphasis on trust and user autonomy is paramount; users must have control over their data, including consent for sensing and data collection. The forthcoming sections will explore the roles of AI as the decision-making 'doorkeeper', jammers as 'physical bodyguards', and RISs as 'magical bodyguards', detailing their contribution to a secure and user-centric ISAC ecosystem.

\subsection{The Doorkeeper: AI-Based Decision Maker}
The AI-based decision maker, or 'doorkeeper', employs advanced machine learning and deep neural networks to control access, authenticate users, detect anomalies, predict threats, and manage resources efficiently, thereby ensuring secure and privacy-conscious communication within the SPPN.

\begin{itemize}
\item {\bf Anomaly Detection and Threat Prediction:} Utilizing AI techniques, the system identifies deviations from normal behavior, flagging potential security or privacy issues in ISAC networks. This capability not only aids in pre-empting security breaches but also optimizes resource allocation by focusing efforts on areas of concern, thus minimizing waste.
\item {\bf Resource Management:} AI optimizes SPPN' resource distribution, including power allocation, on/off keying and beamforming, integrating privacy and security needs into these decisions. Techniques such as federated learning \cite{Shi2022,Anbang2023} offer privacy-preserving machine learning, allowing for collaborative model training without central data consolidation, thus enhancing data security.

\item {\bf User Authentication and Access Control:} AI-driven systems enhance security through accurate user authentication and dynamic access control. By analyzing biometrics and behavior patterns, AI models adapt to ensure only authorized access, adjusting policies based on risk assessments and user profiles, which are continuously refined through machine learning.

\end{itemize}

These AI-driven processes collectively fortify the SPPN, creating a security ecosystem where anomaly detection, resource management, and access control mutually enhance each other. This integration ensures efficient, reliable operation, and a proactive stance on privacy and security, with AI's adaptability playing a central role in maintaining network integrity and user trust.

\subsection{The Physical Bodyguards: Friendly Jammers}

Friendly jamming emerges as a paramount and effective strategy for enhancing physical layer security, serving to obfuscate eavesdroppers and bolster network defenses. This technique entails the creation of sophisticated jamming signals that mimic the protective tactics of an expert bodyguard. By injecting these signals into the eavesdroppers' wireless channels, it becomes challenging for them to decipher confidential information due to a significantly reduced SINR. The primary advantages of friendly jamming include its minimal computational demand and straightforward implementation. Furthermore, it eliminates the need for exchanging coordination messages and additional processing of legitimate signals, streamlining the security process.

This paper presents a case study on the deployment of the friendly jamming framework to safeguard privacy within ISAC networks. Within these networks, numerous authorized transmitters relay confidential data to specific receivers and sensing signals to targeted sensors. The transmitters are positioned randomly, adhering to a homogeneous poisson point process (HPPP). Given the pervasive nature of electromagnetic waves in these networks, legitimate users' private activities risk exposure to unauthorized surveillance. To counteract this vulnerability, we propose the introduction of a third-party service, termed SPPN, which utilizes multiple friendly jammers for protection.

\begin{figure}[t]
\centering
\includegraphics[width=3.6in]{ 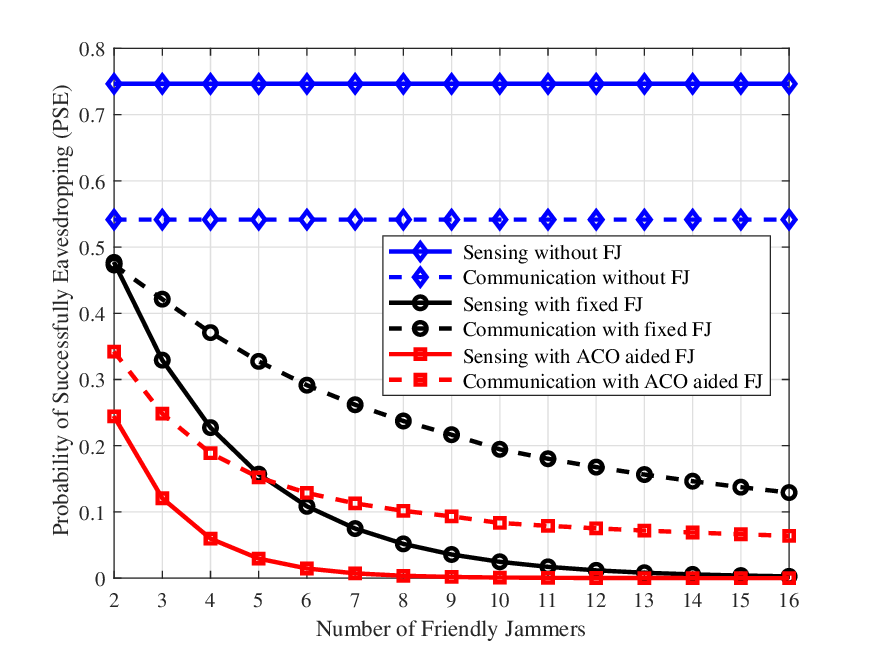}
\caption{PSE versus the number of friendly jammers.}
\label{fig: friendly-jamming_SOP}
\end{figure}

The deployment of friendly jammers within the SPPN is guided by two distinct strategies. Initially, jammers are stationed at predetermined, fixed points. Alternatively, their placement is dynamically optimized through the ant colony optimization (ACO) algorithm, aiming to enhance privacy security while minimizing disruption to legitimate communications. This approach benefits from beamforming technology, directing jamming signals precisely towards potential eavesdropping zones without affecting legitimate communication channels. Except for the friendly jammers, all network devices are equipped with omnidirectional antennas for simplicity.

We evaluate the effectiveness of friendly jamming through the probability of successfully eavesdropping (PSE) metric, which considers an eavesdropping attempt successful if the SINR at the eavesdropper exceeds a specified threshold. Our findings, illustrated in the Fig. \ref{fig: friendly-jamming_SOP}, demonstrate a marked reduction in PSE with the addition of friendly jammers, indicating a substantial decrease in eavesdroppers' ability to intercept private information. Notably, the deployment of more than two jammers significantly diminishes the PSE for both information and sensing channels, with a notable impact on legitimate communications. However, the effectiveness plateaus when the number of jammers exceeds eight, suggesting a diminishing return on further increases in jammer count. The application of the ACO algorithm for dynamic jammer placement further reduces the PSE, significantly enhancing privacy protection with minimal impact on legitimate communications. This highlights the strategic advantage of optimized jammer placement in maintaining robust network security while ensuring efficient communication.

\subsection{The Magic Bodyguards: RISs}
RISs have emerged as a transformative technology, marking the advent of intelligent propagation environments. Through the use of passive metasurfaces equipped with smart controllers, RIS technology allows for the dynamic alteration of electromagnetic wave propagation \cite{Guo2023}. This remarkable ability can be likened to technological alchemy, offering unparalleled precision in directing signal flow. RIS technology is capable of meticulously cancelling out signals prone to interception by eavesdroppers or nullifying harmful transmissions from adversaries, all the while enhancing signal delivery to intended receivers. Such capabilities significantly boost privacy, protect sensitive information, and strengthen wireless communication security. As interest in RIS grows and research advances, it is poised to play a pivotal role in the evolution of secure and privacy-focused ISAC networks, earning it the moniker of 'magic bodyguards' within our proposed SPPN framework.

In practical terms, RIS can be deployed to craft "private zones" or "signal fortresses" within ISAC networks. By precisely managing electromagnetic wave flow, RIS can contain the communication and sensing signals within specified areas, preventing their spread beyond intended boundaries. Furthermore, RIS can establish virtual barriers or selective access zones, thereby improving network access control. RIS also plays a crucial role in setting up secure communication links between trusted entities by optimizing signal paths and selectively boosting or weakening signals. Additionally, RIS supports beamforming techniques that prioritize user privacy by ensuring sensitive information is directed solely at the intended receiver, minimizing the chance of data leakage.

\begin{figure}[t]
\centering
\includegraphics[width=3.6in]{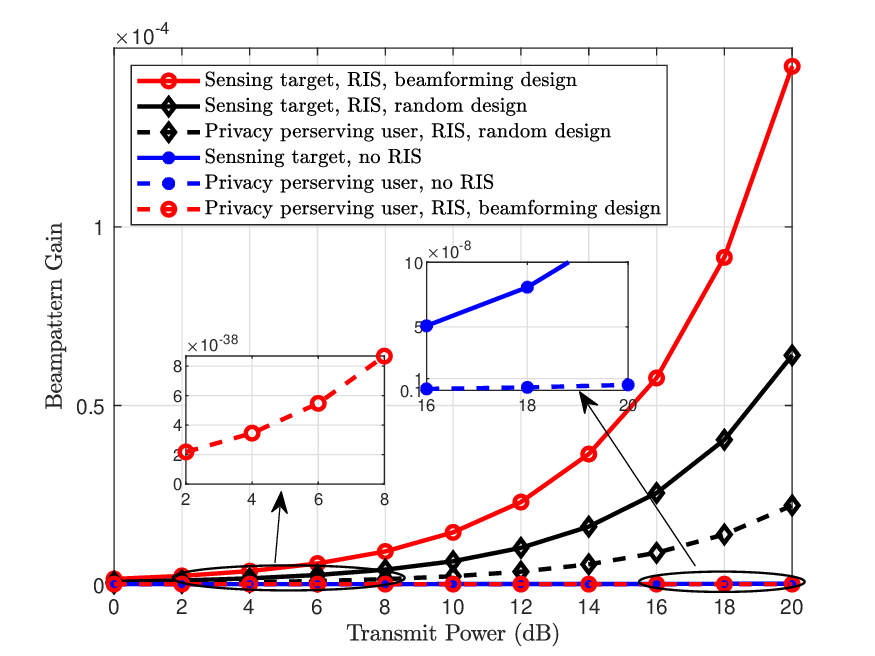}
\caption{Sensing beampattern gain versus signal transmit power.}
\label{fig: RIS_B}
\end{figure}

To demonstrate RIS's potential for user privacy preservation in ISAC networks, we propose an RIS-enhanced solution. Users seeking privacy protection can employ a nearby RIS to obstruct unwanted sensing links from non-trusted entities. The RIS's reflective elements are adaptively tuned to minimize undesirable sensing connections, effectively establishing "private zones". Simultaneously, links to preferred sensing targets are strengthened to increase sensing accuracy. Our analysis, illustrated in Fig. \ref{fig: RIS_B}, shows that with the assistance of an RIS comprising 64 reflective elements, the sensing beampattern gain between the ISAC transmitter and a privacy-conscious user is significantly reduced, nearing zero, while the gain towards a desired sensing target is markedly higher. This disparity confirms the user's preference for non-detection, whereas the desired sensing target remains detectable, underscoring the need for sophisticated reflecting beamforming algorithms to prevent unintended sensing improvements by RISs.

\subsection{How Does SPPN Deal With the Threats?}

 The SPPN framework introduces a robust solution to the challenges outlined in Section II by integrating friendly jammers, RISs, and AI-driven control mechanisms within ISAC networks. Managed by a trusted third-party entity — a consortium that may include a diverse array of stakeholders such as legitimate users, service providers, and government officials — the SPPN collaborates with ISAC service providers to ensure the QoS while simultaneously safeguarding users' privacy.

This collaborative approach enables the SPPN to effectively counteract \emph{Threat 1} by preventing the unnecessary or unauthorized collection of legitimate users' information by ISAC service providers. Furthermore, it enhances network surveillance capabilities against potential security breaches by monitoring non-legitimate users, such as eavesdroppers and attackers. This proactive monitoring assists in safeguarding against information leakage during sensing activities, unauthorized sensing attempts, and mitigates the impact of malicious activities on ISAC QoS, addressing \emph{Threats 2-4}.

Additionally, the SPPN's preventative measures against the overcollection of data from legitimate users at the source significantly reduce the risk of security and privacy breaches within ISAC data centers. By curtailing the amount of data collected to only what is necessary and permitted, the SPPN directly tackles \emph{Threat 5}, minimizing the potential for data misuse and enhancing the overall security and privacy posture of ISAC networks. Through these strategic interventions, the SPPN framework not only preserves the integrity of the ISAC ecosystem but also establishes a foundational layer of trust and security, essential for the advancement of privacy-preserving technologies in the 6G era.

\subsection{Future Research Directions}
The SPPN framework presents a promising avenue for advancing privacy and security in ISAC networks, yet it opens up a spectrum of research challenges and opportunities. Among these, the integration of AI for enhanced network management and control stands out as a critical area for exploration. AI's reliance on extensive and varied datasets for effective training and its continuous adaptation necessitates a careful approach to personal and sensitive data handling to mitigate privacy risks. Furthermore, the opaque nature of AI, especially with deep learning models, poses significant challenges in terms of transparency and explainability. This is particularly problematic in sensitive applications where understanding and justifying security and privacy decisions is paramount.

Concurrently, the development of sophisticated control algorithms for friendly jammers and RISs demands attention. These algorithms must not only ensure privacy and security but also consider the energy efficiency and overall performance of the network. The task is compounded by the difficulties in acquiring accurate channel state information and providing prompt feedback to friendly jammers and RIS, a task made even more daunting by the passive nature of RIS's reflecting elements  and the mobility of targets or users. Additionally, the physical security of RIS against sabotage or destruction by malicious actors poses a significant concern. In environments where RIS are publicly accessible and lack stringent access controls, there is a heightened risk of adversaries compromising RIS control systems or manipulating the surfaces to interfere with communications or siphon off sensitive information.

Addressing these challenges calls for a multidisciplinary approach that spans across the realms of cybersecurity, data privacy, AI, and wireless communications. This entails not only the refinement of AI models for greater transparency and accountability but also the development of resilient control mechanisms for RIS that are robust against physical and cyber threats. Furthermore, innovative strategies for safeguarding RIS against unauthorized access and ensuring the integrity of their operations are imperative. As such, future research must navigate these complex issues, paving the way for the realization of secure, privacy-preserving, and efficient ISAC networks. Through collaborative efforts and continued innovation, the full potential of the SPPN framework can be unlocked, offering a more secure and privacy-focused future for ISAC networks.

\section{Conclusion}

In this paper, we have explored the significant privacy and security challenges associated with ISAC networks due to its open operational environment. We identified key vulnerabilities and technical barriers to achieving a secure and private ISAC ecosystem.To address these concerns, we introduced the SPPN framework, designed to protect against threats and overcome technical complexities. SPPN incorporates AI, friendly jammers, and RISs to bolster network defenses and enhance ISAC's operational capabilities.
Our analysis and case studies highlight SPPN's effectiveness in improving network security and sensing and communication capabilities. The implementation of SPPN marks a step toward a future where ISAC networks operate securely and efficiently, supporting the development of smart applications while safeguarding data privacy and integrity.

\color{black}
\bibliographystyle{IEEEtran} 
\bibliography{IEEEabrv,bib}
\section*{Biographies}
\footnotesize{
\noindent {{KAIQIAN QU}} (qukaiqian@mail.sdu.edu.cn) was born in  Shandong Province, China. He received the B.E. degree from the School of Physics, Zhengzhou University, China, in 2022. Now, he is currently pursuing the M.S.  egree at Shandong University, Jinan, China. His main research interests include the integrated sensing and communication and extremely large-scale MIMO.\\

\noindent{{JIA YE}} (yejiaft@163.com) was born in Chongqing, China. She received the Ph.D. degree in electrical and computer engineering
 from the King Abdullah University of Science and Technology (KAUST), Saudi Arabia, in 2022.  Since January 2023, she has served as a faculty member in the School of Electrical Engineering, Chongqing University, Chongqing, China.  Her main 
research interest includes the performance analysis and modeling of wireless information and energy transfer systems.\\

\noindent{{XURAN LI}} (sdnulxr@sdnu.edu.cn) is currently a lecturer in School of Physics and Electronics, Shandong Normal University (SDNU), China. He received his M.Sc. degree and Ph.D. degree from the Faculty of Information and Technology of Macau University of Science and Technology (MUST) in 2015 and 2018, respectively. His research interests include wireless communication networks, security of wireless networks, and the integrated sensing and communication.\\

\noindent{{SHUAISHUAI GUO}} (shuaishuai$\_$guo@sdu.edu.cn) received the B.E and Ph.D. degrees in communication and information systems from the School of Information Science and Engineering, Shandong University, Jinan, China, in 2011 and 2017, respectively. He visited University of Tennessee at Chattanooga (UTC), USA, from 2016 to 2017. He worked as a postdoctoral research fellow at King Abdullah University of Science and Technology (KAUST), Saudi Arabia from 2017 to 2019. He iscurrently
a Full Professor with Shandong University. His
research interests include 6G communications and machine learning.}










\end{document}